\documentclass[
 aip, apl,
 amsmath,amssymb,
 preprint%
]{revtex4-1}

\usepackage{scalerel,stackengine}
\usepackage{graphicx}
\usepackage{dcolumn}
\usepackage{bm}
\usepackage{color}
\usepackage{soul}
\usepackage{sidecap}
\usepackage{amsmath}
\usepackage{capt-of}

\begin{document}

\title{Quantum phase transition in  ferroelectric-paraelectric heterostructures}

\author{Prasanna Venkatesan Ravindran}
\email[]{pvr@gatech.edu}
\affiliation{
School of Electrical and Computer Engineering, Georgia Institute of Technology, Atlanta, GA 30332, USA
}

\author{Asif Islam Khan}
\affiliation{
School of Electrical and Computer Engineering, Georgia Institute of Technology, Atlanta, GA 30332, USA
}
\affiliation{
School of Materials Science and  Engineering, Georgia Institute of Technology, Atlanta, GA 30332, USA
}
\begin{abstract} 
\noindent 
 
Phase transition between ferroelectricity and quantum paraelectricity via non-thermal tuning parameters can lead to quantum critical behavior and associated emergent phenomena. Ferroelectric quantum critical systems are, however, rare despite the abundance of ferroelectric materials. Here, we show theoretically that in ferroelectric-paraelectric heterostructures, it is plausible to  induce quantum paraelectricity where the quantum temperature ($i.e.$, the temperature below with the onset of ferroelectricity is suppressed by quantum fluctuations) can be tuned by the thickness ratio. This, in turn, can effect a quantum phase transition between effective ferroelectric and quantum paraelectric states, using the thickness ratio as the tuning parameter. The associated quantum critical region offers unexpected prospects in the field of ferroelectric quantum criticality. 

\end{abstract}

\maketitle

Ferroelectric materials are, in general, known to obey the Curie-Weiss law, relating the dielectric constant $\epsilon$ and the temperature $T$, given as
\begin{equation}
    \epsilon = \frac{M}{T- T_C}\label{CW}
\end{equation} 
\noindent where $M$ is the Curie constant, and $T_C$ is the  temperature for the paraelectric-ferroelectric phase transition, $i.e.$, the Curie temperature. Barrett, in 1952\cite{barrett1952dielectric}, suggested a modification to this law by extending Slater's statistical treatment of ferroelectricity\cite{slater1950lorentz} to include quantum mechanical effects. This leads to a quantum statistical relation between $\epsilon$ and $T$, known as the Barrett's formula, which is given as
\begin{equation}
    \epsilon = \frac{M}{\frac{1}{2}T_1\coth\big(\frac{T1}{2T}\big) - T_C}\label{barrett}
\end{equation} 
\noindent where $T_1$ is a quantum temperature, given by $T_1=\hbar\omega/k_B$, with $\hbar$,   $k_B$ and $\hbar\omega$ being the reduced Planck constant,  the Boltzmann constant and twice the zero point energy of the harmonic part of the potential energy of the dipole, respectively. When $T_C$ is much larger than $T_1$, Barrett's formula  reduces to the Curie-Weiss law (Eq. \ref{CW}). On the other hand, if $T_C<T_1$, the material exhibits non-Curie-Weiss behavior. More importantly, if $T_C<T_1/2$, the function: $\epsilon(T)$ lacks a singularity, indicating that there is no paraelectric-ferroelectric phase transition in the system and that the material behaves as a paraelectric in the entire temperature range. The suppression of ferroelectricity below $T_C$ occurs due to the dominance of quantum fluctuations (zero-point motion)  over their thermal counterparts, and hence the phenomenon is often referred to as quantum paraelectricity.

\begin{figure}
    \centering
    \includegraphics[width=0.7\textwidth]{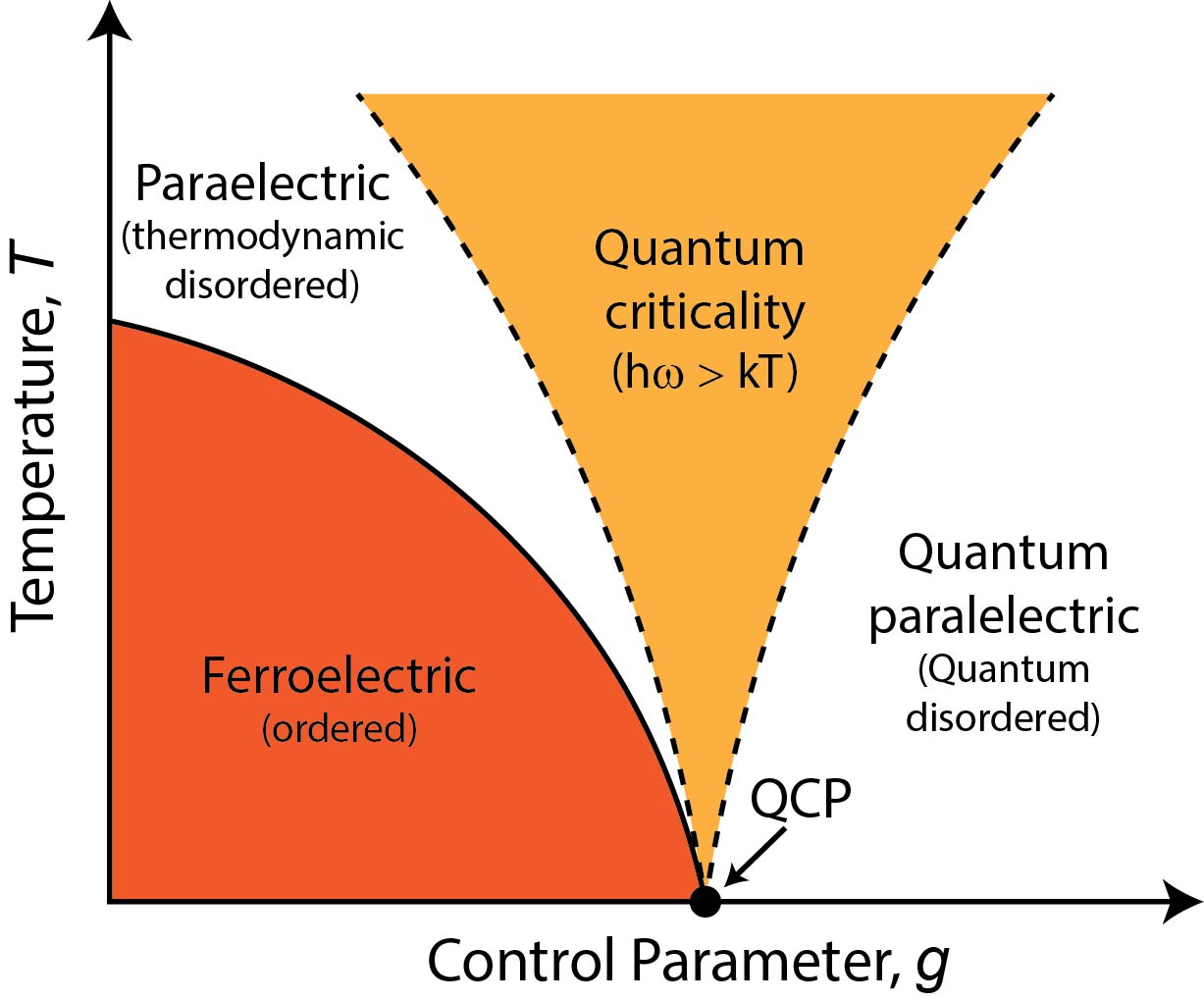}
    \caption{\textbf{Temperature-Control Parameter Phase diagram:} Phase diagram at the ferroelectric-quantum paraelectric phase transition. At 0 K, where thermally no phase transition is possible. But tuning a physical parameter could facilitate quantum fluctuations to suppress the ferroelectric order causing  a phase transition from ferroelectric to a quantum paraelectric phase, giving rise to a quantum critical point. Counter intuitively, the effect of quantum fluctuations extends to larger ranges of the control parameter as temperature increases, leading to a quantum critical region where $\hbar \omega>kT$. This region is filled with exciting effects ranging from superconductivity to polar metallicity.\cite{volkov2020multiband,gegenwart2008quantum}}\label{fig1}
\end{figure}

Recent interests in quantum paraelectricity lie in the context of quantum phase transition and quantum criticality.\cite{rowley2014ferroelectric, narayan2019multiferroic, chandra2017prospects} In the limit of zero temperature where thermal fluctuations are absent,  phase transition with respect to a non-thermal tuning parameter, $g$, such as pressure, composition and so on, can  be driven purely by quantum fluctuations. Quantum phase transitions, as they are referred to\cite{sachdev2007quantum}, are most commonly observed in magnets\cite{lohneysen2007fermi,gegenwart2008quantum,volkov2020multiband}, superconductors and cold atomic/2-D bosonic gases.\cite{endres2012higgs, ranccon2013quantum} The associated quantum critical point, separating the quantum phases,  lies on the 0 K line in the $T$ vs. $g$ phase diagram --- however, the influence of quantum phase transition extends above 0 K into the finite temperatures forming the quantum  critical region (Fig. \ref{fig1}). In this region, the interplay between quantum and thermal fluctuations can lead to novel, unexpected and exotic phenomena, and phases of matter. The most well-known examples of quantum critical behavior are the emergence of unconventional superconductivity\cite{gegenwart2008quantum} and `strange' metallicity.\cite{hayes2021superconductivity, sachdev2011quantum}

Composition\cite{rischau2017ferroelectric}, isotopic substitution\cite{itoh2000quantum} and pressure\cite{coak2019pressure} have been explored as the  tuning parameters to access the ferroelectric quantum critical region in SrTiO$_3$ (STO), the earliest and the most well-known quantum paraelectric.  In fact, ferroelectric quantum criticality has theoretically been shown as the origin of superconductivity in  doped variants of STO.\cite{edge2015quantum} Furthermore, spin can be added to quantum paraelectrics as an additional degree of freedom leading to multiferroics that  exhibit a rich quantum critical behavior around the quantum critical point.\cite{narayan2019multiferroic}   All told,  ferroelectric quantum critical systems are rare. This is despite that ferroelectrics are one of the largest classes of functional materials that includes perovskite oxides,\cite{lines2001principles} organic polymers\cite{bune1998two}, fluorite-structure oxides\cite{boscke2011phase} and layered and two-dimensional materials\cite{liu2016room}, and so on, and that quite a few quantum paraelectric materials are known to exist (see supplementary table S1 for the relative values of $T_1/2$ and $T_C$ for different materials). The rarity of ferroelectric quantum critical system is due to the fact that for quantum criticality, a facile tuning parameter is required  that can tune the relative values of $T_1/2$ and $T_C$ --- which may not be generally available for a wide range of materials.

To that effect, ferroelectric--paraelectric heterostructures, especially in superlattice forms, have been a rich and diverse playground to tune phase transitions --- by mediating the interplay between strain, electrostatics and interfacial effects through careful heterostructure design and epitaxial growth with single atomic layer precision.\cite{lee2005strong, bousquet2008improper, tenne2006probing,  dawber2007tailoring, yadav2016observation, yadav2019spatially, das2019observation, das2021local, li2021subterahertz} Dramatic control of functional properties, such as polarization, capacitance, Curie temperature and so on by superlattice period and thickness ratio has been demonstrated in these structures. Recent observation of intricate  superstructures containing nanometric polarization textures, namely polar vortices\cite{yadav2016observation} and skyrmions\cite{das2019observation}, and functional features therein, such as high-frequency collective responses\cite{li2015sub}, chiral and toroidal order\cite{behera2021electric, damodaran2017phase}, local  static negative dielectric permittivity\cite{yadav2019spatially, das2021local}, and so on has spring-boarded renewed interests in these systems.

In this letter, we discuss the possibility of utilizing heterostructuring to access the quantum critical region, in otherwise ferroelectric materials by tuning the effective Curie temperature. Using a combination of analytical and numerical approaches, we show that it is plausible to engineer quantum paraelectricity in ferroelectric-paraelectric heterostructures. In doing so,  quantum critical behavior can emerge in such heterostructures, where the thickness ratio between the paraelectric and the ferroelectric layers is the tuning parameter.

\begin{figure}
    \centering
    \includegraphics[width=0.7\textwidth]{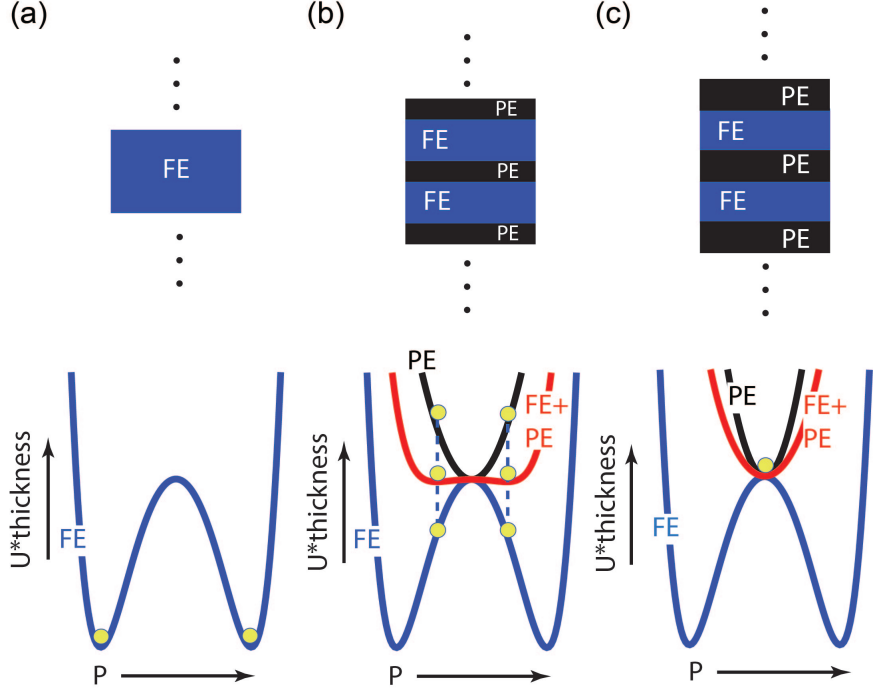}
    \caption{\textbf{(a) Free energy of a stand-alone ferroelectric layer:} The double well energy landscape of a ferroelectric with $T_C>T_1/2$ at $T>T_C$. \textbf{(b)Free energy of a FE-PE heterostructure with an effective ferroelectric phase:} the positive curvature at $P=0$ due to the paraelectric is less than the magnitude of the negative curvature due to the ferroelectric phase. \textbf{(c)Free energy of a FE-PE heterostructure with an effective paraelectric phase:} the positive curvature at $P=0$ due to the paraelectric is more than the magnitude of the negative curvature due to the ferroelectric phase}
    \label{fig_dw}
\end{figure}

To begin with, the Gibbs free energy per unit volume of a ferroelectric material, $U_F$, can be phenomenologically expressed as an even order polynomial of the order parameter $P$, given as  
\begin{equation}
    U_F=\frac{1}{2\epsilon(T)}P^2+ \sum \limits_{i=2,...} \alpha_i P^{2\cdot i}\label{U_F}
\end{equation}

\noindent where $\alpha_i$ are anisotropy constants, temperature independent, and considered positive in this work. In the standard Landau phenomenology of ferroelectricity,  $\epsilon$ is given by Eq. \ref{CW}. To consider  quantum effects, it will be given by Eq. \ref{barrett} in this work. If $T_C>T_1/2$, the energy profile --- the $U_F$ vs. $P$ curve --- of the ferroelectric has a double-well shape at $T<T_C$, as shown in Fig. \ref{fig_dw}(a). We make an assumption, for the sake of simplicity, with regards to a ferroelectric-paraelectric heterostructure that the polarization  is restricted to a direction perpendicular  to the interfaces and is spatially homogeneous. A paraelectric is characterized by a single well energy profile $U_P=P^2/2\epsilon_P$, $U_P$ and $\epsilon_P$ being the free energy per unit volume and dielectric permittivity of the paraelectric material, respectively. The free energy per unit area of the FE-PE structure is a linear combination of the free energy densities of individual layers: $t_FU_F(P)+t_PU_P(P)$, where $t_F$ and $t_P$  are the thicknesses of the ferroelectric layer and the paraelectric layer, respectively. At a given temperature, the functional response of the FE-PE heterostructure can be tuned by the relative values of $t_F$ and $t_P$, as shown in Fig. \ref{fig_dw}. If the curvature of $t_PU_P(P)$ profile is smaller than the magnitude of the negative curvature of the $t_FU_F(P)$ profile at $P=0$, the energy profile,  $U(P)$, retains a double-well shape, resulting in a functional ferroelectric response (Fig. \ref{fig_dw}(b)). On the other hand, if the curvature of $t_PU_P(P)$ profile is larger than the magnitude of the negative curvature of the $t_FU_F(P)$ profile at $P=0$, the energy profile,  $U(P)$, has a single-well shape, leading to a functional paraelectric response (Fig. \ref{fig_dw}(c)).   The free energy density per unit volume of the ferroelectric-paraelectric heterostructure, $U$, is given as

\begin{eqnarray}
    U&=&\frac{t_FU_F+t_PU_P}{t_F+t_P}=\frac{1}{2\epsilon'(T)}P^2+ \sum \limits_{i=2,...} \alpha_i P^{2\cdot i}\label{U}
\end{eqnarray}
\noindent where $\epsilon'$ is the effective dielectric constant of the FE-PE layer. $\epsilon'$ has the same functional form as that in Eq. \ref{barrett}, except that $T_C$ is replaced by an effective Curie temperature $T_C'$, given as 

\begin{equation}
    T_C'=T_C-\frac{M}{\epsilon_P}r  \label{Tc_}
\end{equation}

\noindent where, $r$ is the ratio between the thicknesses of the paraelectric and the ferroelectric layers ($r={t_P}/{t_F}$). According to Eq. \ref{Tc_}, the effect of adding a paraelectric material to or increasing its fraction in a ferroelectric heterostructure is to cause an apparent reduction of the Curie temperature. The temperature at which the paraelectric-ferroelectric transition, $T_{P\leftrightarrow F}$,  occurs can be calculated by finding the pole of 1/$\epsilon'(T)$, which is given as

\begin{equation}
    T_{P\leftrightarrow F} =\frac{T_1}{2\coth^{-1}\frac{2T_C'(r)}{T_1}}\label{P2F}
\end{equation} 
\noindent Note in Eq. \ref{P2F} that if $T_C'>>T_1/2$, $T_{P\leftrightarrow F}$ and $T_C'$ coincides. On the other hand, by setting $T_{P\leftrightarrow F}$=0, one gets the condition for which ferroelectricity is suppressed at 0 K in the ferroelectric-paraelectric heterostructure, making its functional response equivalent to that of a quantum paraelectric.  The critical thickness ratio, $r_C$, above which the FE-PE heterostructure transition from being a ferroelectric to a quantum paraelectric ($i.e.$, $T_C'<T_1/2$) is found to be

\begin{equation}
    r_C =\frac{\epsilon_P}{M}\bigg(T_C - \frac{T_1}{2}\bigg)
\end{equation}

\noindent $r=r_C$ represents a quantum critical point between the ferroelectric and the quantum paraelectric phase in the ferroelectric-paraelectric structure.

To obtain quantitative values of the phenomenological parameters in Eq. \ref{U_F}, we fit the expression to the numerical solution of quantum statistical model of the ferroelectric, which is specified by $T_1$, $T_C$ and $M$. In this quantum model, similar to the procedure presented in Barrett, \cite{barrett1952dielectric} it is assumed that in a single unit cell of the displacement-type ferroelectric, only the ion at the body center moves and that it moves only along the vertical axis, allowing the unit cell to be treated as an anharmonic oscillator. The potential energy of the displaced body-center ion under electric field $E$, is given by $\phi(x) = ax^2 + bx^4 - qxE$. The partition function $z$ of a single oscillator is: $z=\sum_{n=1}^{\infty} \exp(-\varepsilon_n/k_BT)$, where $\varepsilon_n$ is the $n$-th quantized energy level of the oscillator. The partition function for a system of $N$ oscillators, $Z = z^N/N!$. The equivalent Helmholtz free energy in terms of the electric field is given by $A_E = -k_BT\log Z$. The ionic polarizability of the body-center ion is $\alpha' = (\partial P/\partial E)/N$, with $P$ being the order parameter: polarization, given as  $P = -(\partial A_E/\partial E)_T$. The free energy of the ferroelectric material with independent body-center ions, $A_P$ calculated as $A_P = A_E + PE$, can be expressed as $A_P = \alpha'P^2 + \beta'P^4$. The interaction of ions with each other is taken into account using Lorentz correction.\cite{slater1950lorentz} For Lorentz correction constants, $c_3$ and $c_4$, the Gibbs free energy per unit volume of the ferroelectric, $A_P= \alpha P^2 + \beta P^4$, with $\alpha$ and $\beta$ being given by

\begin{align}
    \alpha &= c_3^2\alpha' + \frac{c_3c_4}{2\epsilon_0}\\
    \beta &= c_3^3\beta'
\end{align}

\noindent For a perovskite crystal structure, $c_3$ and $c_4$ were calculated by Slater.\cite{slater1950lorentz} The procedure to calculate the microscopic parameters $a$ and $b$, and the Lorentz correction terms $c_3$ and $c_4$ for given values of $T_1$, $T_C$ and $M$ is detailed in the supplementary section I.

For numerical calculations, the material parameters for the heterostructures are as follows: $T_1/2$=25 K, $T_C$=390 K, $M$=1.5x10$^5$ K $\times\epsilon_\circ$, and $\epsilon_P$=200$\epsilon_\circ$, where $\epsilon_\circ$ is the vacuum permittivity. The microscopic parameters ($a$, $b$,  $c_3$ and $c_4$) calculated for these values are listed in supplementary table S1. Fig. \ref{fig3}(a) and \ref{fig3}(b) plot 1/$\epsilon'$ and the remanent polarization $P_\circ$ as functions of $T$  of the ferroelectric-paraelectric heterostructure for different values of $r$. The shift of $T_C'$ with an increasing $r$ is observed. For $r=$ 0.5, 0.52 and 0.7, 1/$\epsilon'$ vs. $T$ curves flatten out for $T<T_1/2$, and $P_\circ$ is zero at all temperatures. Fig. \ref{fig3}(c) and \ref{fig3}(d) show  phase plots of  $P_\circ$ and $1/\epsilon'$ in the ($T$, $r$) plane, respectively. In Fig. \ref{fig3}(c), the boundary between $P_\circ \ne 0$ and $P_\circ = 0$ regions represents the $T_{P\leftrightarrow F}$ contour. At the zero-temperature limit ($T$=0 K), the ferroelectric to paraelectric transition occurs at $r=0.49$, which represents the quantum critical point, $r_C$.

Fig. \ref{fig4} plots $T_{P\leftrightarrow F}$ and $T_C'$ as functions of $r$. We observe that $T_{P\leftrightarrow F}$ follows $T_C'$ until $T_C'\approx T_1$, and  $T_{P\leftrightarrow F}=0$ K, when $T_C'= T_1/2$. Note that $T_C'=0$ K at $r=0.52$, and for $r>0.52$, the ferroelectric-paraelectric heterostructure would have exhibited paraelectricity at all temperatures --- even in the classical case ($i.e.$, in the hypothetical case where quantum fluctuations are absent). While quantum effects manifest for all value of $r$ in the form of slowly varying $\epsilon'$ and $P_\circ$ at $T<T_1/2$ (as seen in Fig. \ref{fig3}(a) and \ref{fig3}(b), respectively), it is only in the range  $r \in (r_C, \epsilon_PT_C/M)$ that quantum fluctuations suppress ferroelectricity that would have been present in the classical scenario. 

So far, we considered the paraelectric to have a temperature-independent dielectric permittivity, for the sake of providing an intuitive picture of the quantum phase transition in the system. We also modeled the heterostructures where the paraelectric has a temperature dependence, $e.g.$ governed by the Barrett formula. Supplementary Fig. S3(a) and S3(b) show the phase plots of $P_\circ$ and $1/\epsilon'$, respectively, in the ($T-r$) plane for a BaTiO$_3$(BTO)-SrTiO$_3$(STO) heterostructure. We also varied the parameters of the ferroelectric and the paraelectric to understand their impact on the phase diagram, a few examples of which are shown in Fig. S3(c-f). In all cases, we observe that the phase diagrams show  clear boundaries between the ferroelectric, the paraelectric  and the quantum paraelectric phases and the quantum critical points.

By varying the quantum temperature of the paraelectric, $T_{1,P}$, while keeping that of the ferroelectric, $T_{1,F}$, a constant for different value of $r$, we observe that  the effective $T_1$ in a ferroelectric-paraelectric heterostructure, $T_{1,{FP}}$, lies  between  $T_{1,F}$ and $T_{1,P}$ (Fig. \ref{fig:T1}). This indicates that in addition to the effective Curie temperature ($T_C'$), the quantum temperature of the ferroelectric-paraelectric heterostructure is also tunable by the thickness ratio.

\noindent

It is encouraging to note that, experimentally, tens to hundreds of K of shift in the effective Curie temperature with the change of the individual layer thicknesses has been observed in PbTiO$_3$/SrTiO$_3$, BaTiO$_3$/SrTiO$_3$ and PbSrTiO$_3$/SrTiO$_3$ superlattices.\cite{bousquet2008improper, tenne2006probing,  dawber2007tailoring, Zubko2010xray, Zubko2012Ferroelectric, Zubko2012Ferroelectric, zubko2016negative} This indicates that  classical phase transition occurs in these systems where  thickness ratio is the tuning parameter.  Electric displacement in such structures is spatially non-uniform, leading to intricate polarization textures, such as 180$^\circ$ domains\cite{Zubko2010xray}, polar vortices\cite{yadav2016observation} and skyrmions\cite{das2019observation} --- all of which are also strongly affected by the epitaxial strain. Admittedly, studies into these exciting phenomena are, however, limited to the classical regime, away from the quantum limits in the cryogenic temperature range. These effects, which are not considered in our theoretical analysis, can add richness to the predicted quantum critical region in ferroelectric-paraelectric heterostructures.

In summary, we have discussed the possibility of a quantum paraelectric phase in ferroelectric heterostructures. Further we have also presented the possibility of achieving tunable quantum paraelectricity in ferroelectric-paraelectric heterostructure, where the quantum temperature, the temperature below which the onset of an effective ferroelectricity is suppressed due to  quantum fluctuations, and the transition temperature from ferroelectric to quantum paraelectric phase --- can be tuned  by the thickness ratio. This raises the prospect of observing quantum phase transition and a quantum critical region in such systems, with the thickness ratio as the tuning parameter.   Experiments geared at elucidating if a quantum phase transition is indeed present in the zero-temperature limit in these heterostructures and at extracting the scaling laws that characterize the ferroelectric quantum critical regime are an exciting avenue for future research.

This work was supported by the Georgia Tech Quantum Alliance (GTQA). The authors thank Jayakanth Ravichandran and Martin Mourigal for fruitful discussions. 

\newpage
\bibliographystyle{apsrev4-1}

\bibliography{ms}
\newpage

\newpage

\begin{figure}
    \centering
    \includegraphics[width=0.7\textwidth]{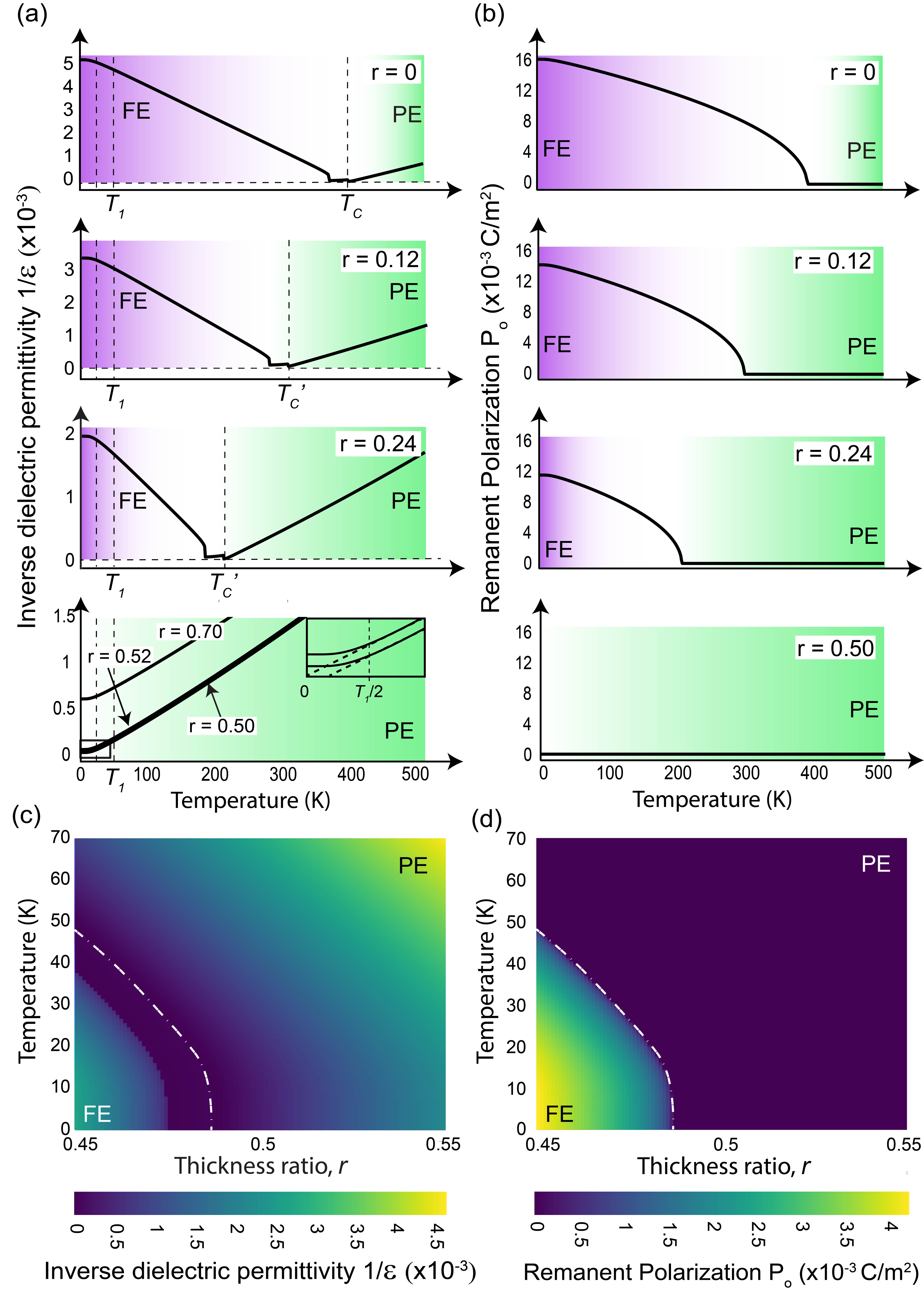}
    \caption{\textbf{(a) Inverse dielectric permittivity and (b) Remanent Polarization:} Variation of the inverse dielectric permittivity and remanent polarization with temperature is shown for different values of thickness ratio illustrating the decrease in transition temperature with increasing $r$. When $r>r_C$, $T_C'$ reduces below $T_1/2$ and the heterostructure becomes a quantum paraelectric as shown in the inset in (a) and the zero-remanent polarization. For $r>\epsilon_PT_C/M$, the heterostructure becomes a paraelectric with $T_C'<0$. \textbf{(c\&d) Ferroelectric-Paraelectric Phase Diagrams:} Temperature-thickness ratio phase diagrams show the (c) inverse dielectric permittivity and (b) remanent polarization of the heterostructure. The white line represents the phase boundary between the ferroelectric and paraelectric phases. }
    \label{fig3}
    \vspace{-0.2in}
\end{figure}

\begin{figure}
     \centering
     \includegraphics[width=0.7\textwidth]{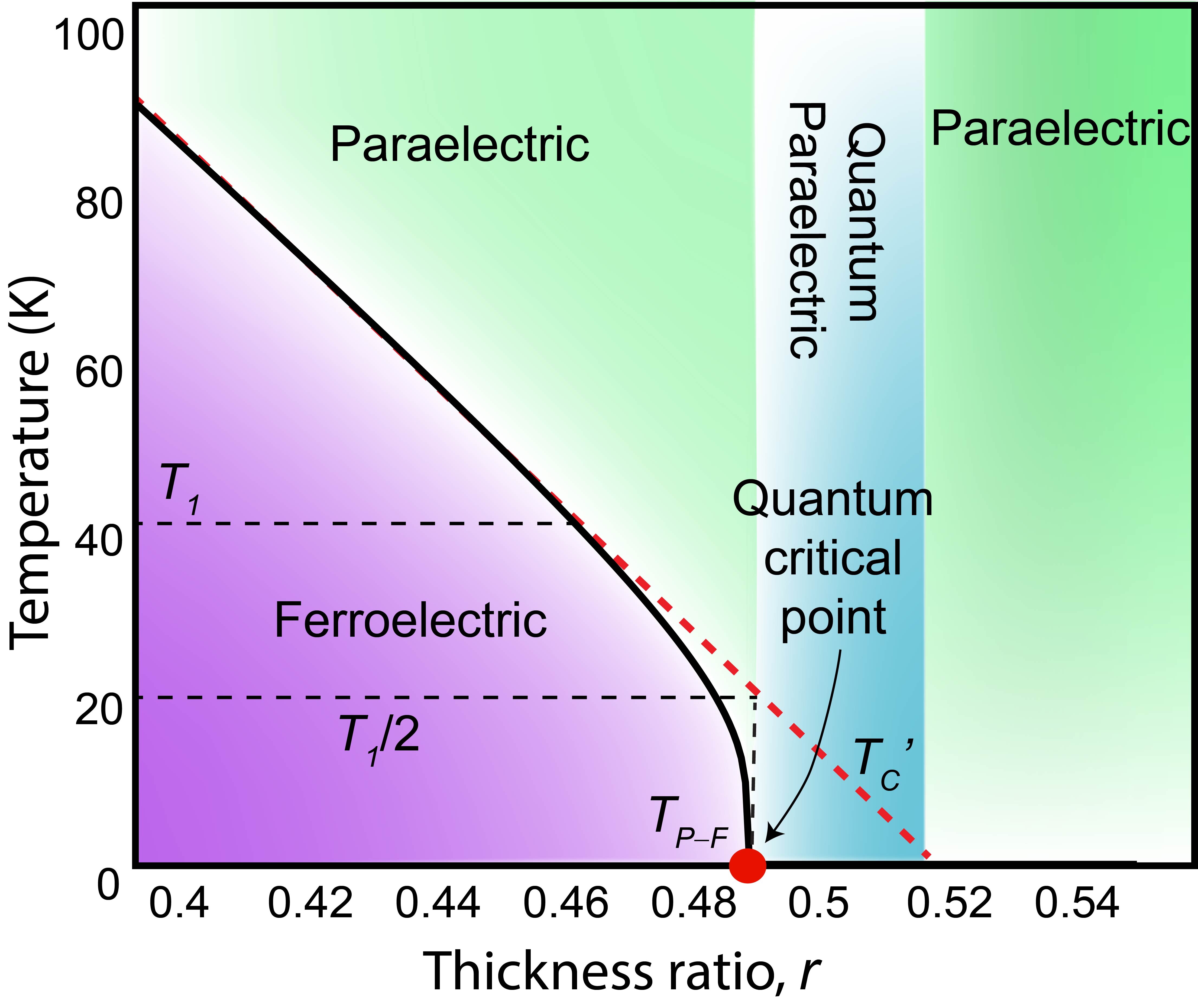}
     \caption{\textbf{Temperature-Thickness ratio phase diagram:} The T-r plane shows the phase transition from ferroelectric to quantum paraelectric phase. The phase transition is tuned by the ratio of thickness of the paraelectric and ferroelectric layers, $r$. At 0 K, the heterostructure transitions from a ferroelectric to quantum paraelectric phase at $r=r_C$, the quantum critical point. The system remains in this incipient ferroelectric phase for $r$ ranging from $r_C$ to $\epsilon_PT_C/M$. For $r>\epsilon_PT_C/M$, the heterostructure acts as a paraelectric for any temperature as predicted by the classical theory. }
     \label{fig4}
     \vspace{-0.2in}
 \end{figure}

\begin{figure}
    \centering
    \includegraphics[width=0.7\textwidth]{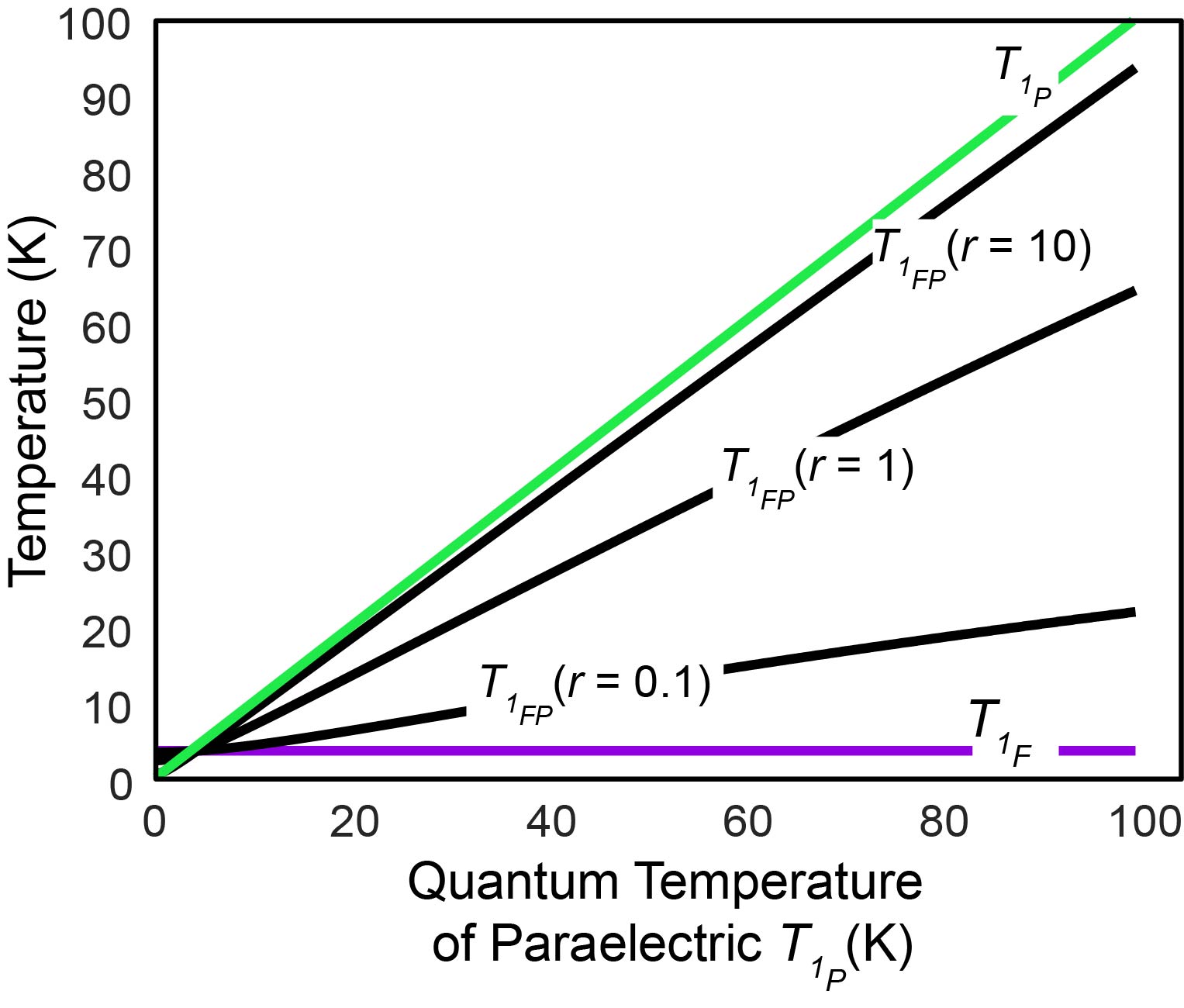}
    \caption{The effective quantum temperature $T_{1_{FP}}$ of the FE-PE heterostructure with the ferroelectric as BTO and a paraelectric layer with $M_P$ = 1.55x10$^5$x$\epsilon_\circ$ K and $T_{1_{P}}$ varying from 0.1 K to 100 K is shown for different thickness ratios.}    \label{fig:T1}
\end{figure}

\end{document}


\title{Quantum phase transition in  ferroelectric-paraelectric heterostructures}

\author{Prasanna Venkatesan Ravindran}
\email[]{pvr@gatech.edu}
\affiliation{
School of Electrical and Computer Engineering, Georgia Institute of Technology, Atlanta, GA 30332, USA
}
\author{Asif Islam Khan}
\affiliation{
School of Electrical and Computer Engineering, Georgia Institute of Technology, Atlanta, GA 30332, USA
}
\affiliation{
School of Materials Science and  Engineering, Georgia Institute of Technology, Atlanta, GA 30332, USA
}

\maketitle

\newpage

\section{Microscopic parameters and Lorentz parameters}
The polarization from a single unit cell is given described by the spring constants, $a$ and $b$ and the Lorentz parameters, $c_3$ and $c_4$. For a ferroelectric whose macroscopic dielectric permittivity is given by Eq. 2 can be described by the macroscopic parameters, $M$, $T_1$ and $T_C$. All these parameters are related by the following equations:\cite{slater1950lorentz,barrett1952dielectric}

\begin{align}
    T_1 &= \frac{\hbar \nu}{k}\\
    \nu &= \sqrt{\frac{2a}{m}}\\
    T_0 &= \frac{2a^3 \epsilon}{3Nq^2bkD}\bigg(\frac{Dq^2N}{2a\epsilon} - 1\bigg)\\
    M &= \frac{2a^3 \epsilon B}{3Nq^2bkD}\\
    b &\approx a/10^{-20}\\
    B &= -\frac{1}{c_3 c_4}\\
    D &= -\frac{c_4}{c_3}
\end{align}

\begin{figure*}[t]
    \centering
    \includegraphics[width=0.9\textwidth]{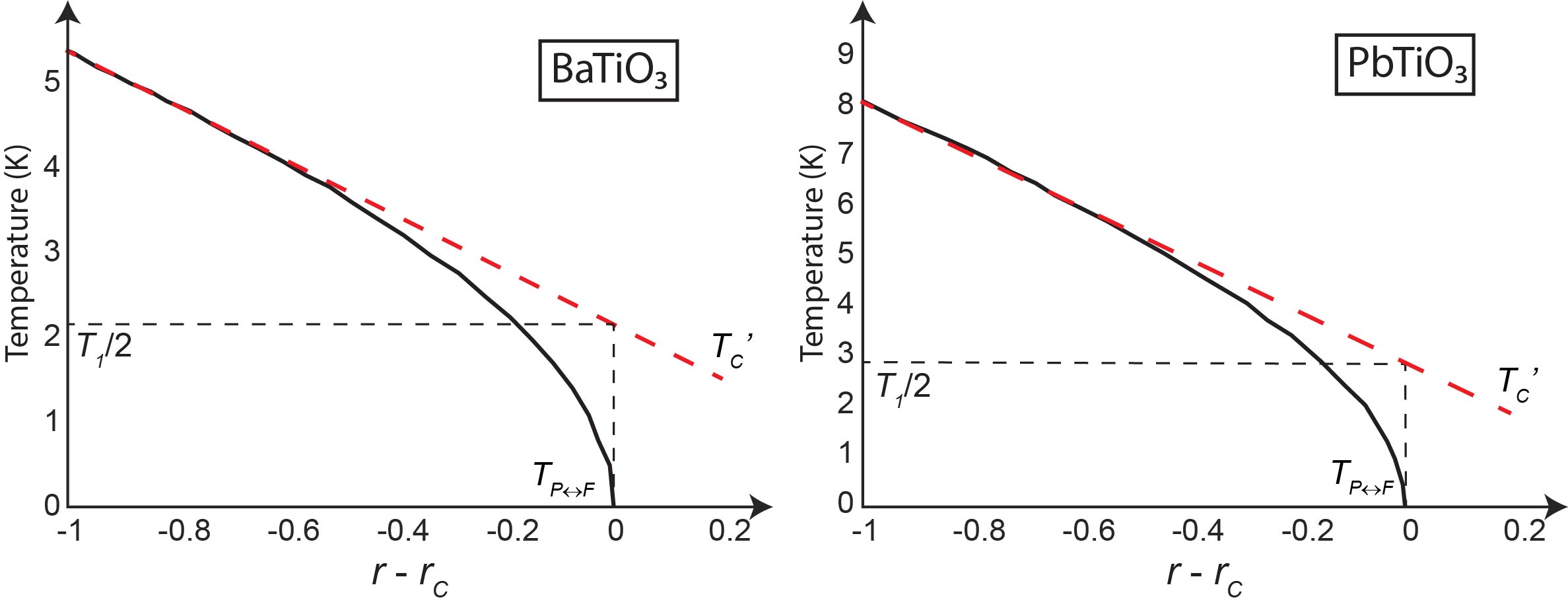}
    \caption{ Variation of the transition temperature $T_{P\leftrightarrow F}$ as a function of pressure for (a) BaTiO$_3$ and (b) PbTiO$_3$ is shown by the solid black curve from Fig. 8 in Ref.\cite{roussev2003theory}. The x-axis is the pressure represented as $r-r_C$, where $r_C$ is the pressure at the quantum critical point of the phase transition. The red dashed line shows the variation of effective Curie temperature $T_C'$ with pressure. the dielectric permittivity of the paraelectric to be independent of the temperature, the effective Curie temperature when $r=r_C$ is the half of the quantum temperature $T_1/2$.}
    \label{t1}
\end{figure*}

The microscopic parameters and the Lorentz factors for a few ferroelectrics are shown in the table 1. The $T_1$ value for BaTiO$_3$ and PbTiO$_3$ was calculated from the Fig. \ref{t1} in ref. \cite{roussev2003theory}. Assuming the dielectric permittivity of the paraelectric to be independent of the temperature, $T_1$ is the effective Curie temperature of the material for which $r-r_C=0$. Table \ref{classification} shows the classification of the materials based on the relative values of $T_1/2$ and $T_C$.

\begin{table}[b]
\caption{Microscopic parameters and Lorentz correction factors of displacive ferroelectrics. FE$_1$ is the model ferroelectric used for the numerical calculations in the manuscript.}
\centering
\begin{ruledtabular}
\begin{tabular}{ccccccccc}
Material & $T_1$/2 (K) & $T_C$ (K) & $M/\epsilon_0$ (K) & m (kg) & $a$ (N/m) & $c_3$ & $c_4$\\  \hline
KTaO$_3$\cite{fujishita2016quantum}      & 25.67   & 11.8 & 6x10$^4$ & 3.00x10$^{-22}$ & 6.7648x10$^3$ & 0.3019 & -90.1971\\
SrTiO$_3$\cite{fujishita2016quantum}     & 33.1     & 24.5  & 1.55x10$^5$ & 7.95x10$^{-23}$ & 2.9806x10$^3$ & 0.1878 & -24.7259 \\ 
EuTiO$_3$\cite{katsufuji2001coupling}     & 81     & -25  & 2.34x10$^4$ & 7.95x10$^{-23}$ & 1.7849x10$^4$ & 0.4835 & -381.0874 \\
NaMnF$_3$\cite{dubrovin2020incipient}     & 80     & -7  & 4.35x10$^3$ & 9.12x10$^{-23}$ & 1.9974x10$^4$ & 1.1214 & -989.0665 \\
BaFe$_{12}$O$_{19}$\cite{shen2016quantum}     & 23.65     & -22.9  & - & - & - & - & -  \\
BaTiO$_3$\cite{roussev2003theory,barrett1952dielectric}     & 2.2      & 390  & 1.5x10$^5$ & 7.95x10$^{-23}$ & 1.3167x10$^1$ & 0.1675 & -0.1110 \\
PbTiO$_3$ \cite{haun1987thermodynamic,roussev2003theory,haun1987thermodynamic}    & 2.9  & 765  & 1.5x10$^5$ & 7.95x10$^{-23}$ & 2.28795x10$^1$ & 0.1645 & -0.1929\\
FE$_1$     & 25   & 390 & 1.5x10$^5$ & 8.00x10$^{-23}$ & 1.711x10$^3$ & 0.1908 & -14.4285
\end{tabular}\label{table}
\end{ruledtabular}
\end{table}

\begin{table}[b]
\caption{Classification of materials based on their transition temperatures}
\centering
\begin{ruledtabular}
\begin{tabular}{ccc}
Material & Relation between $T_C$ and $T_1$ & Examples\\  \hline
Ferroelectric & $T_C>T_1/2$ & BaTiO$_3$, PbTiO$_3$, etc.\\
Quantum paraelectric & $0<T_C<T_1/2$ & SrTiO$_3$, KTaO$_3$, etc.\\
Paraelectric & $T_C<0$ & EuTiO$_3$, NaMnF$_3$, etc.\\
\end{tabular}\label{classification}
\end{ruledtabular}
\end{table}

\section{Quantum-Statistical model}

\begin{figure*}[h]
    \centering
    \includegraphics[width=0.9\textwidth]{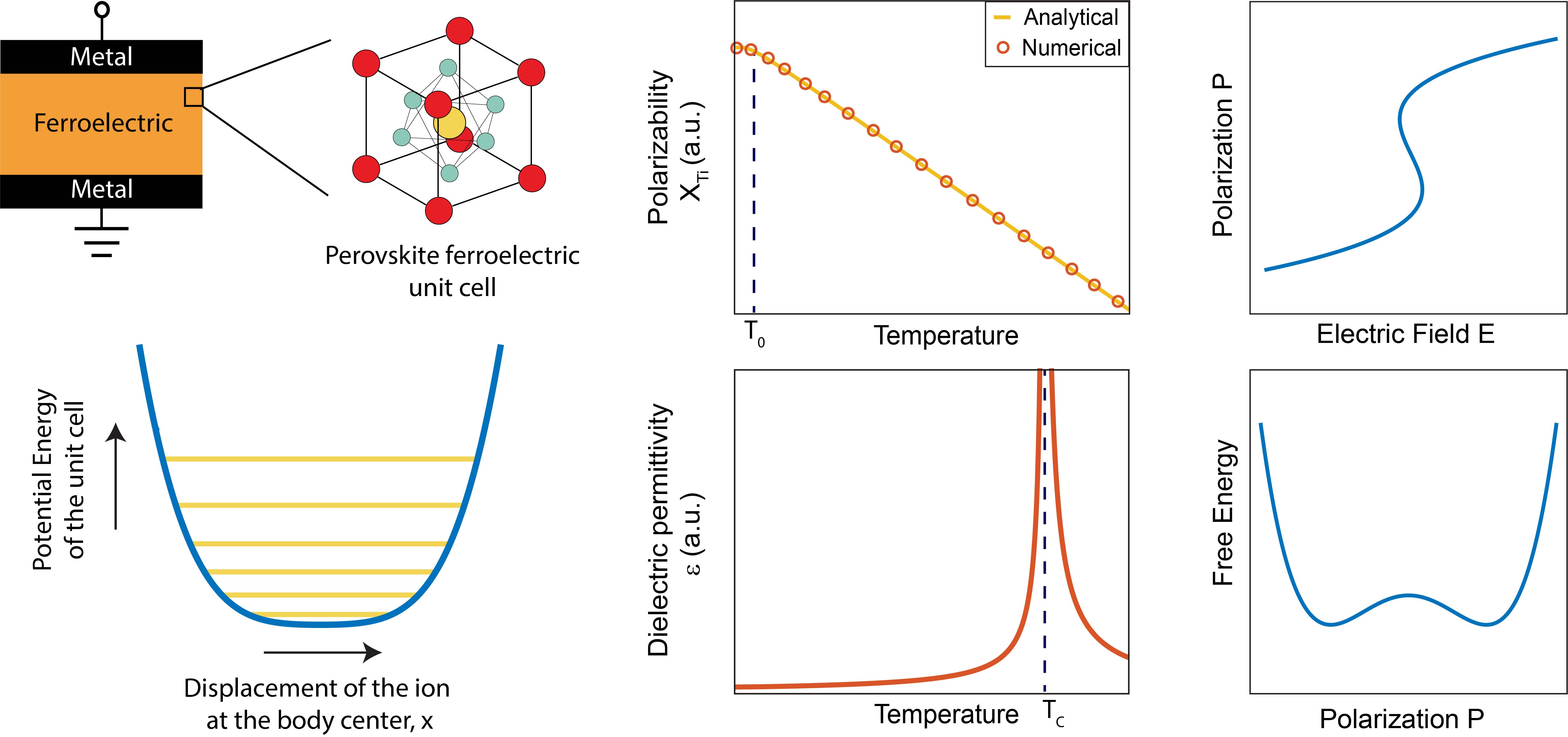}
    \caption{(a) MFM structure and body-center cubic perovskite unit cell. (b) Anharmonic potential energy of the body-center ion as a function of the displacement $x$ with quantized energy levels.(representative) (c) Ionic polarizability from numerical calculations overlaid on Eq. 2.\cite{barrett1952dielectric} (d) Temperature-dependence of dielectric permittivity with phase transition at $T_C$. (e\&f) P-E curve and free energy landscape at $T<T_C$}
    \label{fig:model}
\end{figure*}

A displacive ferroelectric is modeled as follows:
\begin{enumerate}
    \item Hamiltonian of a unit cell modeled as an anharmonic oscillator(Fig. \ref{fig:model}) is given by:
    \begin{equation}
        \phi(x) = ax^2 + bx^4 - qxE
    \end{equation}
    \item The Schrodinger's equation for the anharmonic oscillator is solved to obtain the quantized energy levels:
    \begin{equation}
        \phi\psi = \epsilon\psi
    \end{equation}
    \item The partition function $z$ for a single unit cell is given by:
    \begin{equation}
        z=\sum_{n=1}^{\infty} \exp(-\varepsilon_n/k_BT)
    \end{equation}
    \item The partition function is extended to N unit cells giving the partition function of the system:
    \begin{equation}
        Z = z^N/N!
    \end{equation}
    \item From the partition function of the system, the equivalent Helmholtz free energy $A_E$ is calculated:
    \begin{equation}
        A_E = -k_BT\log Z
    \end{equation}
    \item The order parameter, $P$ is calculated by taking the first derivative of the Helmholtz free energy:
    \begin{equation}
        P = -(\partial A_E/\partial E)_T
    \end{equation}
    \item The equivalent Gibbs free energy $A_P$ is calculated from $A_E$ and $P$:
    \begin{equation}
        A_P = A_E + PE
    \end{equation}
    \item $A_P$ as a function of $P$ is obtained by fitting $A_P(E)$ to $\alpha'P(E)^2 + \beta'P(E)^4$:
    \begin{equation}
        A_P(P) = \alpha' P^2 + \beta' P^4
    \end{equation}
    \item The free energy is corrected using the Lorentz correction to account for the contributions from the local electric field:
    \begin{align}
        \alpha &= c_3^2\alpha'P^2 + \frac{c_3c_4}{2\epsilon_0}P^2\notag\\
    \beta &= c_3^3\beta'P^4\notag\\
        A_P(P) &= \alpha P^2 + \beta P^4
    \end{align}
    
\end{enumerate}

\section{Ferroelectric-Paraelectric Heterostructures}

\begin{figure*}[t]
    \centering
    \includegraphics[width=0.9\textwidth]{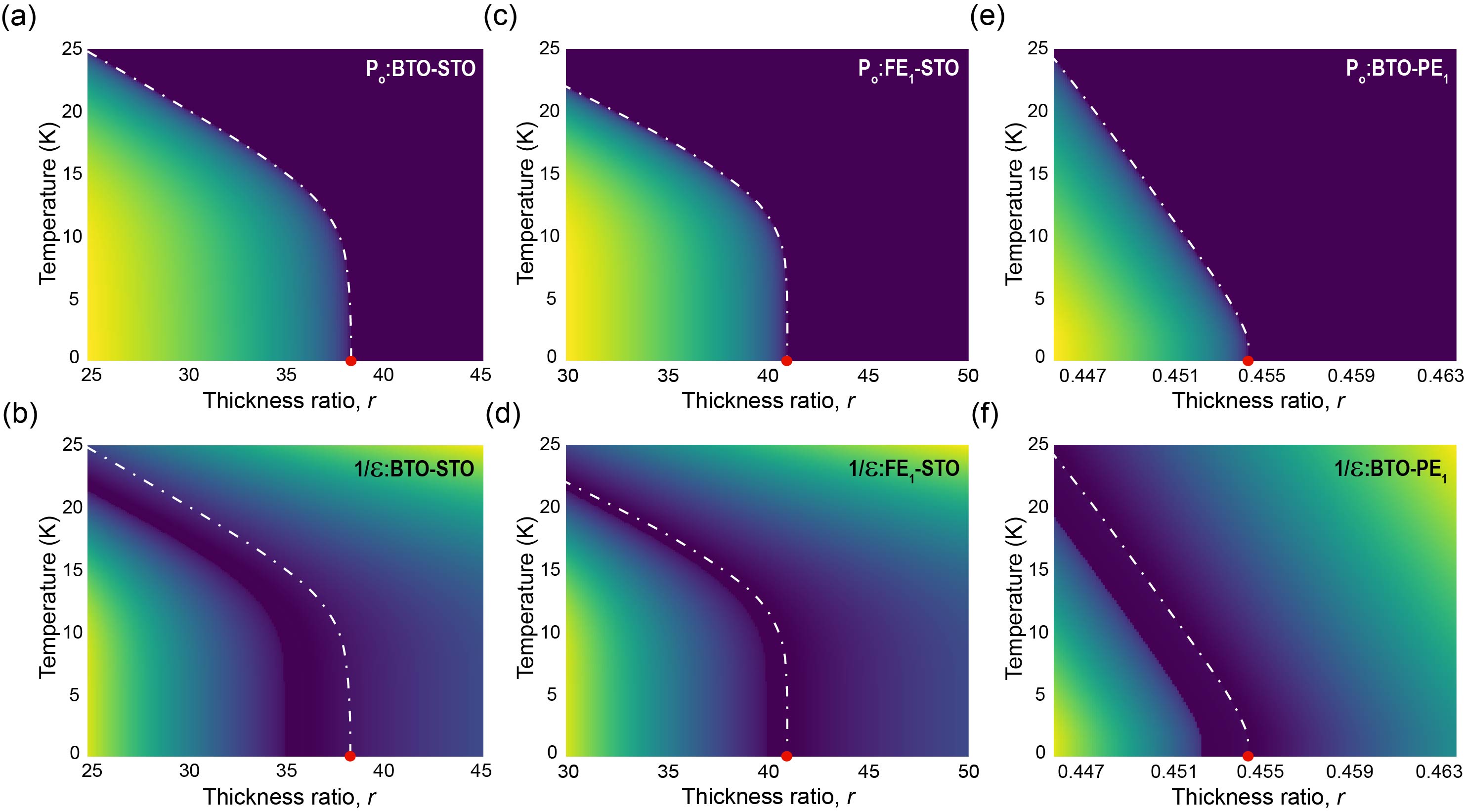}
    \caption{Remanent polarization $P_\circ$ and Inverse dielectric polarization $1/\epsilon$  of (a-b) BTO-STO, (c-d) FE$_1$($T_1$ = 50 K, $T_C$ = 390 K, $M$ = 1.5x$10^5$x$\epsilon_\circ$ K)-STO and (e-f) BTO-PE$_1$($\epsilon_P = 200$). The red dot indicates the quantum critical point $i.e.$ thickness ratio $r=r_C$ at which the FE-PE heterostructure transitions from ferroelectric to paraelectric phase at 0 K}
    \label{fig:heterostructures}
\end{figure*}

\noindent Consider a FE-PE heterostructure consisting of the ferroelectric and paraelectric layers described by the parameters [M$_F$, T$_{1_F}$ and T$_{C_F}$] and [M$_P$, T$_{1_P}$ and T$_{C_P}$], respectively. 
\begin{equation}
       \epsilon_P = \frac{M_P}{\frac{T_{1_P}}{2}\coth\bigg(\frac{T_{1_P}}{2T}\bigg) - T_{C_P}}
\end{equation}
\begin{equation}
    \epsilon_F = \frac{M_F}{\frac{T_{1_F}}{2}\coth\bigg(\frac{T_{1_F}}{2T}\bigg) - T_{C_F}}
\end{equation}

\noindent For a thickness ratio $r$, the heterostructure has an effective dielectric permittivity $\epsilon_{FP}$ which can be expressed in terms of Eq. 2 with effective parameters $M_{FP}$, $T_{1_{FP}}$ and $T_{C_{FP}}$ as shown below. 

\begin{equation}
    \frac{1+r}{\epsilon_{FP}} = \frac{1}{\epsilon_{F}} + \frac{r}{\epsilon_{P}}
\end{equation}
\noindent where, 
\begin{equation}
    \epsilon_{FP} = \frac{M_{FP}}{\frac{T_{1_{FP}}}{2}\coth\bigg(\frac{T_{1_{FP}}}{2T}\bigg) - T_{C_{FP}}}
\end{equation}

\noindent The effective Curie temperature $T_C'$ of this FE-PE heterostructure is:

\begin{equation}
    T_C'=\frac{M_P T_{C_F} + r T_{C_P} M_F}{M_P + r M_F}  \label{Tc_}
\end{equation} 

\noindent The critical thickness ratio, $r_C$, above which the FE-PE heterostructure transitions from being a ferroelectric to a quantum paraelectric at 0 K is found to be,

\begin{equation}
    r_C =-\frac{M_P}{M_F}\frac{(2T_{C_F} - T_{1_F})}{(2T_{C_P} - T_{1_P})}
\end{equation}

\noindent Fig. \ref{fig:heterostructures} shows the phase diagram and quantum critical points for three different heterostructures including BTO-STO. The microscopic parameters ($a$, $b$, $c_3$, $c_4$) for BTO are estimated from the first principle \textit{ab initio} calculations of temperature-pressure phase diagrams in Ref. \cite{roussev2003theory}, and for STO, these parameters obtained from experiments reported in Ref. \cite{vendik1997modeling,fujishita2016quantum}. The parameters are listed in supplementary table S1, and the parameter estimation procedure is described in supplementary section I. It can be seen that the effective quantum temperature of BTO-STO is larger than that of BTO, thereby providing better accessibility to quantum critical region.

\bibliography{supplementary_material}